\documentclass[aps,pra,floatfix,twocolumn,superscriptaddress]{revtex4-1}
\usepackage[final]{graphicx}
\usepackage{times,amsmath,amssymb}
\usepackage{epsfig,color}
\usepackage{hyperref}
\usepackage[caption = false]{subfig}
\usepackage{thumbpdf,enumerate}
\usepackage{booktabs}

\definecolor{blu}{RGB}{97, 129, 179}
\definecolor{ocra}{RGB}{223, 155, 48}
\definecolor{verde}{RGB}{141,177, 61}
\definecolor{rosso}{RGB}{234, 95, 55}
\definecolor{viola}{RGB}{136, 119, 177}

\begin{document}

\title{Multiparameter quantum estimation of noisy phase shifts}

\author{Emanuele Roccia}
\affiliation{Dipartimento di Scienze, Universit\`a degli Studi Roma Tre, Via della Vasca Navale 84, 00146, Rome, Italy}

\author{Valeria Cimini}
\affiliation{Dipartimento di Scienze, Universit\`a degli Studi Roma Tre, Via della Vasca Navale 84, 00146, Rome, Italy}

\author{Marco Sbroscia}
\affiliation{Dipartimento di Scienze, Universit\`a degli Studi Roma Tre, Via della Vasca Navale 84, 00146, Rome, Italy}

\author{Ilaria Gianani}
\affiliation{Dipartimento di Scienze, Universit\`a degli Studi Roma Tre, Via della Vasca Navale 84, 00146, Rome, Italy}

\author{Ludovica Ruggiero}
\affiliation{Dipartimento di Scienze, Universit\`a degli Studi Roma Tre, Via della Vasca Navale 84, 00146, Rome, Italy}

\author{Luca Mancino}
\affiliation{Dipartimento di Scienze, Universit\`a degli Studi Roma Tre, Via della Vasca Navale 84, 00146, Rome, Italy}

\author{Marco G. Genoni}
\affiliation{Quantum Technology Lab, Dipartimento di Fisica, Universit\`a degli Studi di Milano, 20133, Milan, Italy}

\author{Maria Antonietta Ricci}
\affiliation{Dipartimento di Scienze, Universit\`a degli Studi Roma Tre, Via della Vasca Navale 84, 00146, Rome, Italy}

\author{Marco Barbieri}
\affiliation{Dipartimento di Scienze, Universit\`a degli Studi Roma Tre, Via della Vasca Navale 84, 00146, Rome, Italy}
\affiliation{Istituto Nazionale di Ottica - CNR, Largo Enrico Fermi 6, 50125, Florence, Italy}

%
%

\begin{abstract}
Phase estimation is the most investigated protocol in quantum metrology, but its performance is affected by the presence of noise, also in the form of imperfect state preparation. Here we discuss how to address this scenario by using a multiparameter approach, in which noise is associated to a parameter to be measured at the same time as the phase. We present an experiment using two-photon states, and apply our setup to investigating optical activity of fructose solutions. Finally, we illustrate the scaling laws of the attainable precisions with the number of photons in the probe state.
\end{abstract}

\maketitle


\section{Introduction}

The identification of genuine properties of a system, from single molecules to a complex composite systems, represents a primary goal for physical and chemical analysis. As  metrological requirements become increasingly demanding in terms of performance, understanding the ultimate precision achievable in the estimation of a parameter represents a key issue. In this respect, quantum metrology, aiming at designing protocols to perform  optimal measurements, figures as the most appealing and intriguing field of research and applications~\cite{giovannetti2004,giovannetti2006}.

Phase estimation has long represented the heart of quantum metrology~\cite{Paris08,giovannetti2011,RafalReview}: in a large number of technological areas the estimation problem is concerned with determining a single parameter, and this is typically manifested as a phase shift of the quantum state describing the probe. The engineering of such state then aims at providing the optimal choice for an enhanced sensitivity in the estimation: particular families of states, as squeezed~\cite{caves1981,breitenbach1997,rozema2014} or {\it N}00{\it N} states~\cite{lee2002,afek2010,giovannetti2011,joo2011}, are often used to feed interferometers, showing how nonclassicality represents the primary ingredient of the probe states. Nevertheless, an increase in sensitivity balances the robustness of the quantum state: the more these resources are informative, the more they are difficult to obtain and fragile. Their metrological yield can then be spoilt by ungoverned or spurious couplings, unavoidable in any real experiment \cite{dorner2009,kacprowicz2010,peter2011,datta2011,escher2011,knysh2011,demko2012,gill2000,macchiavello2003,ballester2004,pinel2013,genoni2013,sergey2013,crowley2014,Birchall2016}

In the context of noisy quantum metrology, several attempts have been done in order to restore a quantum advantage \cite{Matsuzaki2011,Chin12, Chaves2013,Kessler14,Arrad14,Dur14, Brask2015, Sekatski2015a,Plenio2016,Gefen2016, Smirne16, Haase2017, Gorecka2017,Layden2017,Sekatski2017,Matsuzaki2017,Zhou2018,Albarelli2018}; in all these metrological schemes, a proper characterization of the noise affecting the system is however required. It is not always the case that such characterization can be performed in advance: for instance, in time-varying cases the noise process itself can change, and it is then important to design strategies that treat the assessment of both unitary parameters, such as phases, and dissipative parameters, including loss or phase diffusion, at equal pace by demanding a multiparameter approach. Such extended characterization is akin in spirit to channel tomography~\cite{orieux2013,rozema2014,zhou2015}, aside from the important difference that one allows for a single choice of probes, and not for a tomographically complete family. Multiparameter has been the subject of intensive research over the last years, and this has highlighted the emergence of a trade-off in the achievable precision on individual parameters in many practical instances~\cite{vaneph2013,Vidrighin14,pezze2017,magdalena2017,roccia2017a}. On the other hand, working in a multiparameter setting also brings the advantage of making the estimation process more robust against small deviations of the designed probes from the optimal states~\cite{Vidrighin14}. 

Here we present an estimation experiment in which the multiparameter approach is followed to obtain the value of a phase shift and, at the same time, a reliable estimate of the quality of the probe that actually investigates the material, corresponding in our case to the mode indistinguishability of two input Fock states. 
At difference from \cite{Birchall2016} where phase-estimation with not-perfectly-indistinguishable photons is investigated, the resources are devoted to both estimation tasks. Hence, by estimating both quantities at the same time, it is possible to reduce biases due to an uncertain knowledge of the probe. 
We apply this method to the investigation of chiral aqueous solutions of fructose investigated by two-photon  {\it N}00{\it N} states. The theoretical generalization to higher photon numbers $N$ demonstrates the presence of a trade-off in the scaling associated to the precision on phase and mode distinguishablity.

\section{Two-parameter estimation of phase and visibility}

A common setup in quantum phase estimation uses single-photon pairs produced via a spontaneous parametric down conversion (SPDC) process: in the typical scheme, the two photons are first combined on a beam splitter (BS), so that Hong-Ou-Mandel interference~\cite{HOM1987} produces in a $N$00$N$ state with $N{=}2$, {\it i.e.} a state in a superposition of two photons being present in either mode, and none on the other. The monitored element, imparting a phase shift $\phi$, is then inserted on one of the modes, with the other left unperturbed. The detection scheme has the two modes recombined on a second BS, and photon counters on the outputs. The combination of the nonclassicality of the state and of the optimality of the measurement choice results in oscillations of the photon counting probabilities occurring with a phase $2\phi$, hence in a superior precision than attainable with classical light of the same average energy. This strategy, although effective, clashes with the non-ideal visibility $v$ of the two-photon interference on the two BSs: a second characteristic parameter to be estimated is then introduced. The value of $v$ is limited both by the distinguishabilty of the two photons in spectral and spatial degrees of freedom, but also to dephasing or depolarisation mechanisms taking places inside the sample. Therefore a preliminary calibration performed under conditions which do no reflect those present at the time of phase estimation might weaken of the metrological capabilities of the protocol. We have then explored the alternative approach of assessing the values of $\phi$ and $v$ simultaneously.   

\begin{figure}[t]
\centering
{\includegraphics[width=0.95\linewidth]{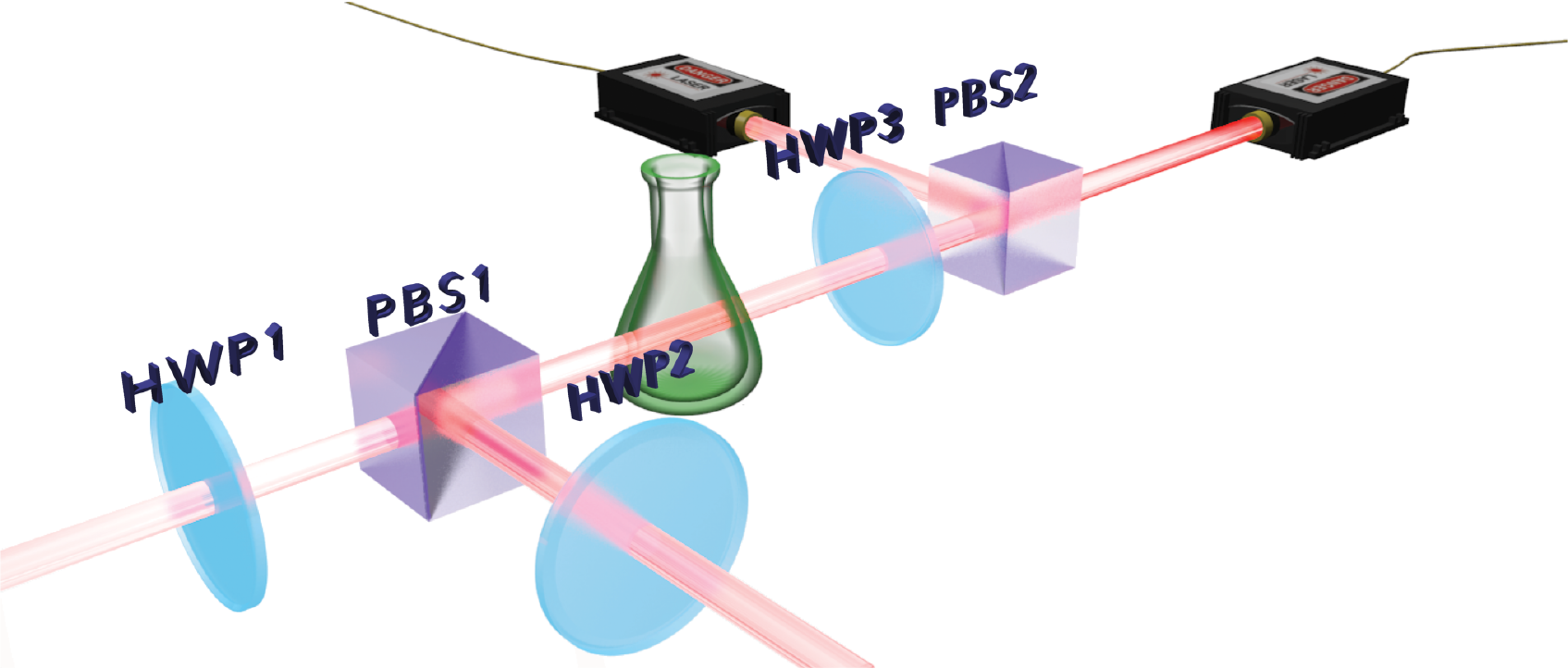}}
\caption{Experimental set-up: each one of the two single photons (wavelength 810nm ) of the pair generated via type-I  SPDC from a Beta Barium Borate (BBO, 3mm-length) non-linear crystal excited via a continuous wave (80mW power) pump laser, passes through an half wave plate (HWP1 at 0$^{\circ}$ and HWP2 at 45$^{\circ}$) before being combined on a polarized beam splitter (PBS1). These photons are used to estimate the birefringent phase imparted by the optical activity of a chiral solution. A wave plate (HWP3) and a second polarizer (PBS2) project the outcoming photons onto different polarizations. In the calibration procedure, an additional HWP, not sketched here, replaces the solution to impart a well-defined phase.}
\label{fig:setup}
\end{figure}

\begin{figure*}[t!]
\centering
\includegraphics[width=0.435\textwidth]{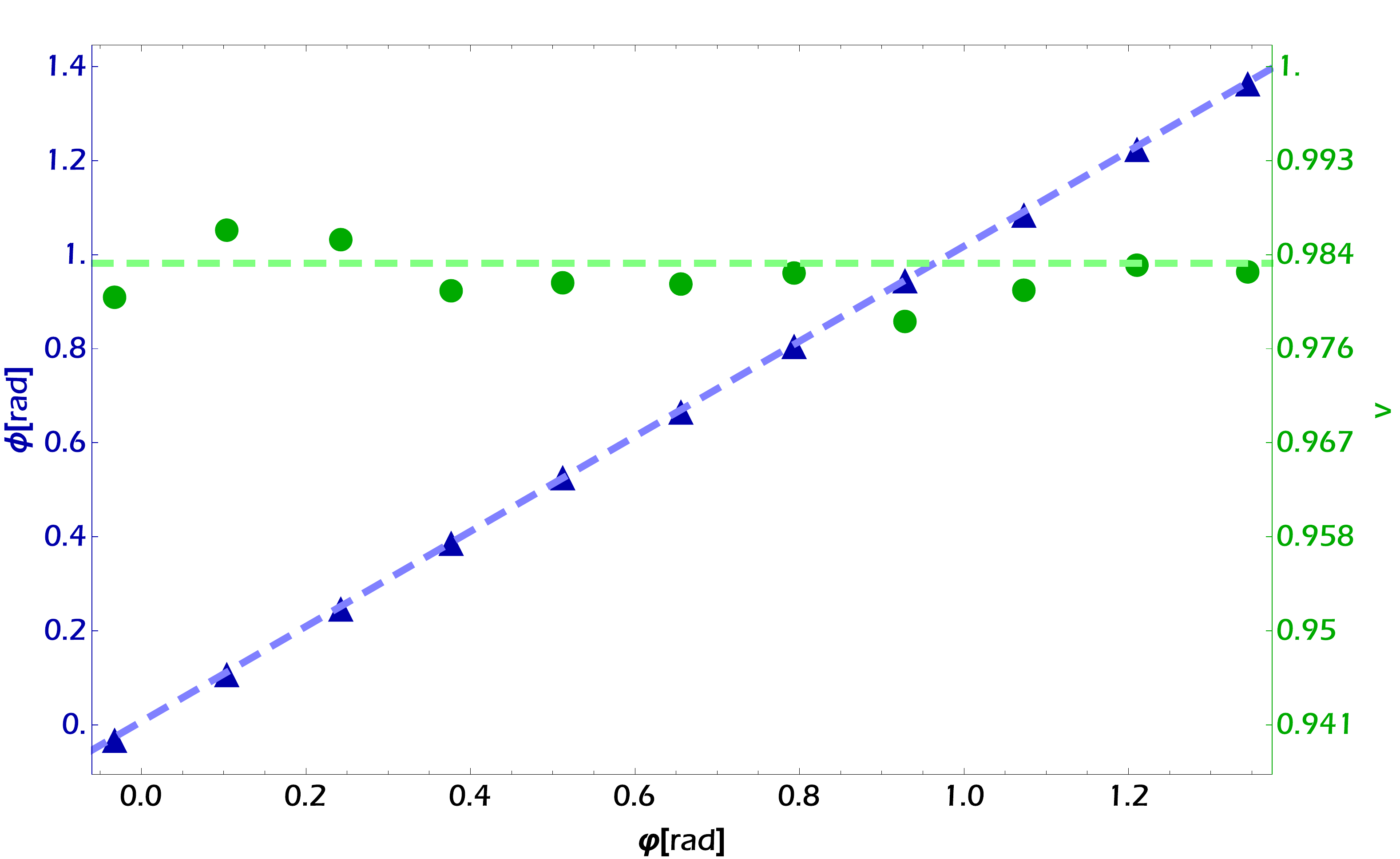}
\includegraphics[width=0.4\textwidth]{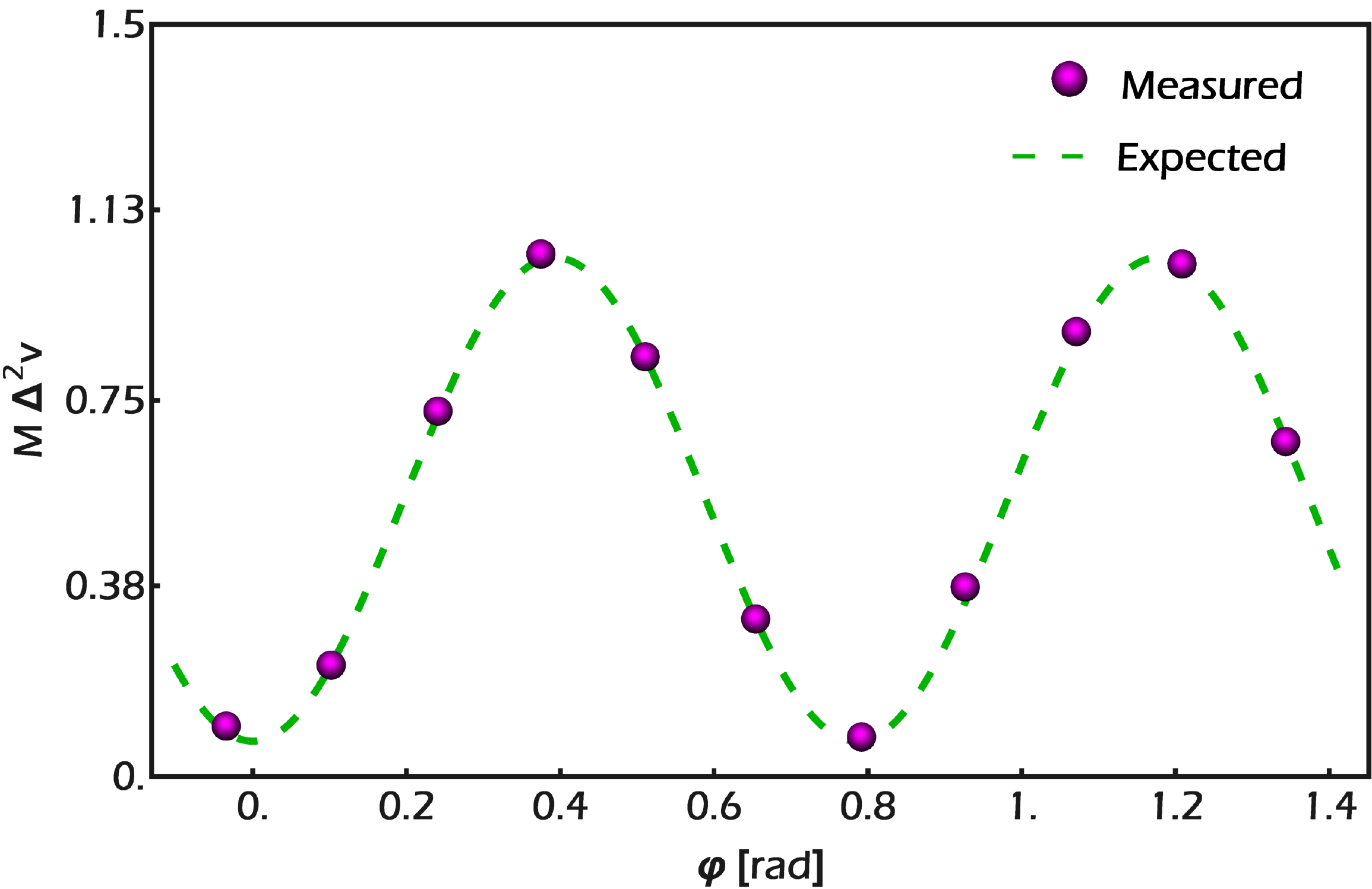}
\includegraphics[width=0.4\textwidth]{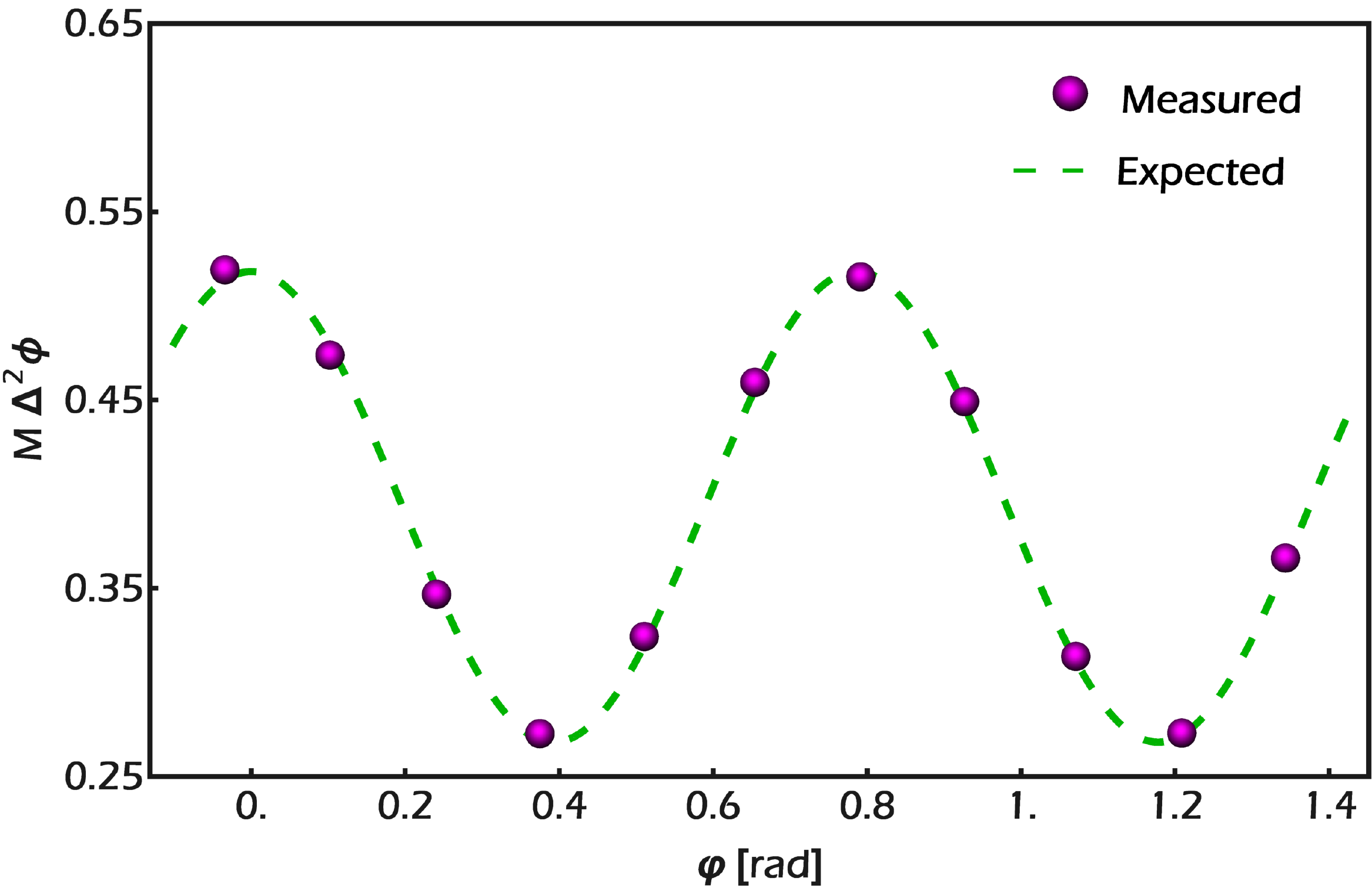}\hspace{0.5cm}
\hspace{0.5cm}\includegraphics[width=0.4\textwidth]{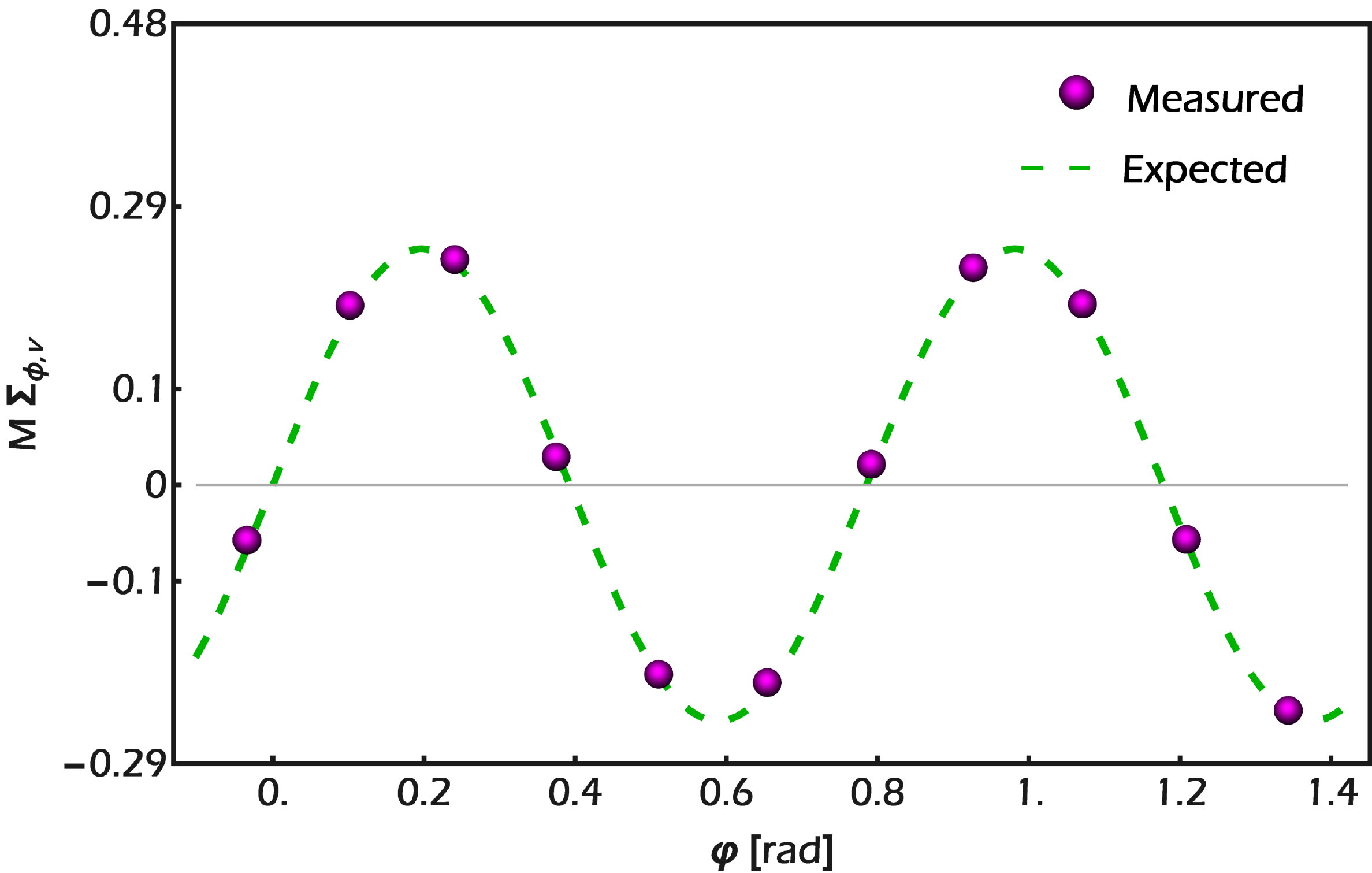}
\caption{Multiparameter Bayesian estimation for setup calibration. Panel (a): estimated phase (blue triangles, left scale) and visibility (green circles; right scale) vs. calibration phase. Dashed lines are linear fit of data. Panel (b) and (c): estimated variance (times the number of resources $M$) for visibility (b) and phase (c) as a function of the imparted phase. The dashed line represents the corresponding CRBs. Panel (d): estimated covariance for the visibility and phase as a function of the imparted phase. The dashed line represents the corresponding CRB. All covariance matrices have been estimated from $M\simeq70$K repetitions. Error bars are smaller than the marker size for all data.}
\label{fig:calibration}
\end{figure*}

Figure~\ref{fig:setup} shows the experimental apparatus we used to implement the phase estimation. Two photons with mutually orthogonal polarizations, horizontal ($H$) and vertical ($V$), are combined on a polarizing beam splitter (PBS). Having very similar spectra, the two photons are highly indistinguishable, and their perfect interference would produce the {\it N}00{\it N} state in the left- ($L$) and right-circular ($R$) polarization modes:
\begin{equation}
\begin{aligned}
\hat{a}^{\dagger}_H \hat{a}^{\dagger}_V \vert0\rangle=&\frac{1}{2}\left(({\hat{a}^{\dagger}_R})^2-({ \hat{a}^{\dagger}_L})^2\right)\vert0\rangle\\
=&\frac{1}{\sqrt{2}}\left(\vert2_R,0_L\rangle-\vert0_R,2_L\rangle\right).
\end{aligned}
\end{equation} 
Introducing a phase $\phi$ on the $R$-mode is equivalent to rotating a linear polarization by an angle $\phi/2$, and modifies the state as
\begin{equation}
\vert \psi \rangle=\cos{\phi}\, \hat{a}^{\dagger}_H \hat{a}^{\dagger}_V \vert0\rangle-\sin{\phi}\,\frac{(\hat{a}^{\dagger}_H)^2-(\hat{a}^{\dagger}_V)^2}{2} \vert0\rangle,
\label{eq:state}
\end{equation} 
The phase $\phi$ modulates the populations in the states $\vert \uparrow \rangle = \vert 1_H,1_V\rangle$ and $\vert \downarrow \rangle = \left(\vert2_H,0_V\rangle-\vert0_H,2_V\rangle\right)/\sqrt{2}$, which represent the basis of an effective two-level system, {\it i.e.} a qubit. The detection scheme consists of a half wave plate (HWP) and a second PBS, allowing to select  arbitrary linear polarizations via the angular position $\theta$ of the HWP. Photon counting is performed by fiber-coupled avalanche photodiodes (APD) place of each of the two output arms from the PBS. In the realistic case when the modulations in the state defined in \eqref{eq:state} occurs with visibility $v$, the relevant detection probabilities are: 
\begin{equation}
\begin{aligned}
p_1(\theta|\phi,v)&=\frac{1}{1+v}\left(1+v\cos(8\theta-2\phi)\right)\\
p_2(\theta|\phi,v)&=\frac{v}{1+v}\sin^2(4\theta-\phi).
\end{aligned}
\label{eqn:prob_2par}
\end{equation}
where $p_1(\theta|\phi,v)$ describe the probability of a coincidence count between the two arms (associated to $\vert \uparrow \rangle$), and $p_2(\theta|\phi,v)$ is the probability of finding two photons in either arm (both events are associated to $\vert \downarrow \rangle$). Because of the underlying single-qubit structure of the state in \eqref{eq:state}, at least two settings of $\theta$ must be chosen to resolve the two parameters: this amounts to performing a positive-operator valued measurement (POVM) with 2$\times$3 outcomes. Furthermore, since our detectors can not resolve the photon number, we have actually adopted four settings of $\theta$ (viz. $\theta=\{0,\;\pi/16,\;\pi/8,\; 3\pi/16\}$), and used the post-selected probabilites
\begin{equation}
p(\theta|\phi,v)=\frac{1}{4}\left(1+v\cos(8\theta-2\phi)\right),
\end{equation}
which only consider the coincidence events for each setting. In the post-selection picture, the probability above treats $\theta$ as the outcome of the measurement scheme. Assuming that the four settings are in fact performed randomly, each one with probability 1/4, Eq. (4) quantifies the probability that the coincidence event detected corresponds to the particular setting $\theta$. Data are collected in the form of a vector $\bar n$, formed by four coincidence count rates $n_\theta$ associated to each setting $\theta=\{0,\;\pi/16,\;\pi/8,\; 3\pi/16\}$: therefore, we post-select 4 out of the possible 4$\times$3 outcomes.

An experimental joint distribution for the measured values of $\phi$ and $v$ is obtained by Bayesian estimation. This consists in using Bayes's theorem to update the {\it a priori} joint probability $P_A(\phi,v)$, based on the knowledge of the measured values $n_\theta$: $P_B(\phi,v | \bar n) = {\mathcal N} P_A(\phi,v)  \prod_\theta p(\theta|\phi,v)^{n_\theta}$ ($ {\mathcal N}$ is a normalization constant). 

\section{Experimental Results}

We have tested the performance of our experiment with a calibration step, by inserting an additional HWP between the two PBSs: this imparts a set phase $\varphi$ depending on its angle setting, and provides of the metrological capabilities of our multiparameter strategy. Fig.~\ref{fig:calibration}a shows, as a function of the imparted phase, the results of the measured of $\phi$ and $v$ from $P_B(\phi,v|\bar n)$, quantified as the first moments of the marginal distributions $\phi_B$ and $v_B$: 
\begin{equation}
\begin{aligned}
&\phi_B=\int \phi\,P_B(\phi,v|\bar n) d\phi\,dv,\\
&v_B=\int v\,P_B(\phi,v|\bar n) d\phi\,dv,
\end{aligned}
\end{equation}
with the integration limits set by the width of $P_A(\phi,v)$. A linear regression of the values highlights the goodness of the phase estimation $\phi$, as its slope is $s_\phi=1.011\pm0.004$, in agreement with the expected value 1. Concerning the visibility, the estimation appears to be affected by fluctuations around a constant mean value instead, as the slope of the linear fit of that data confirms, $s_v=-0.001\pm0.003$. Such fluctuations can be consider as the manifestation of spurious effects not accounted for in our modelling, such as different optical coupling of the initial $H$ and $V$ photons onto the two different fibers.

\begin{figure*}[t]
\centering

\includegraphics[width=0.47\textwidth]{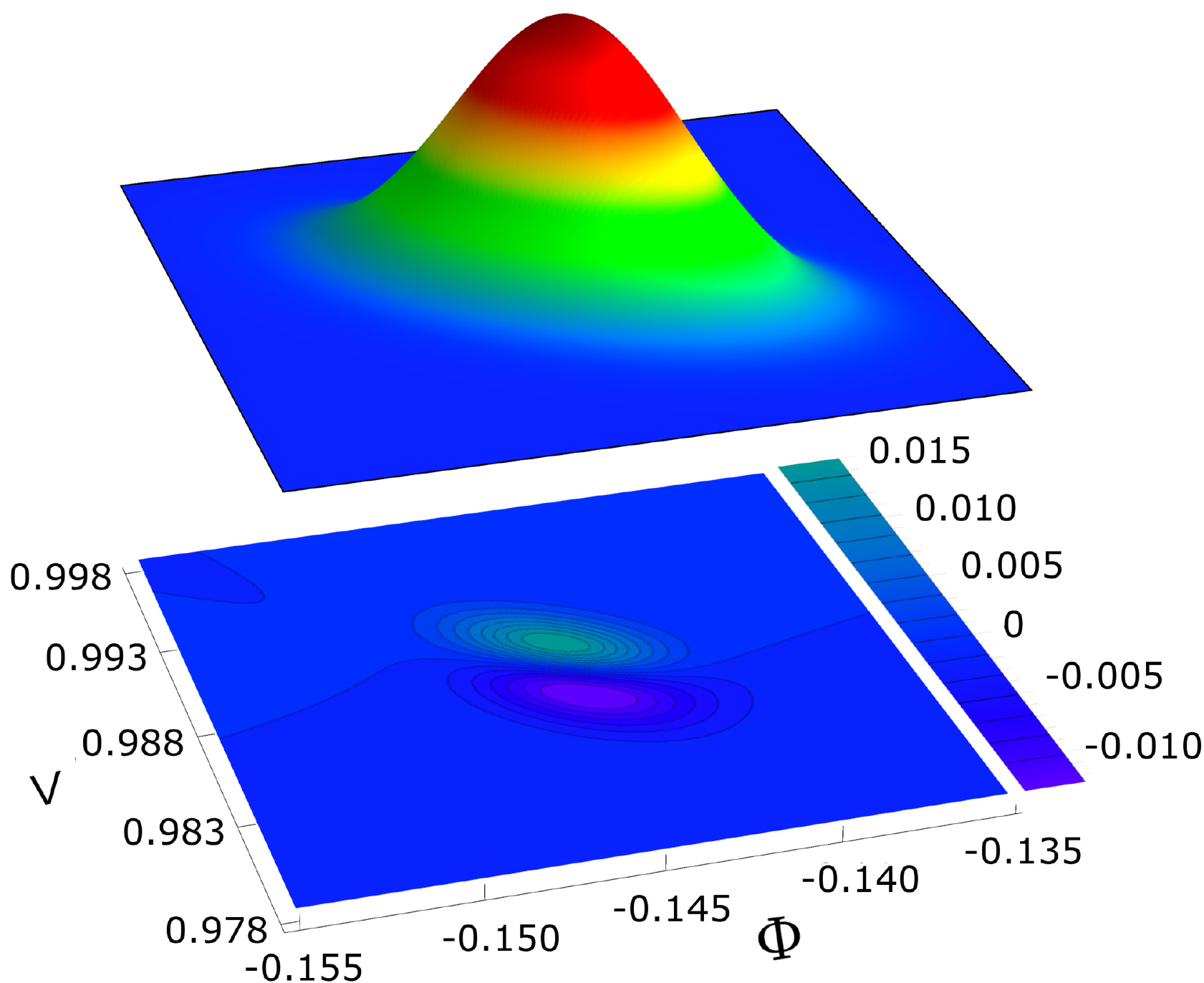}
\includegraphics[width=0.47\textwidth]{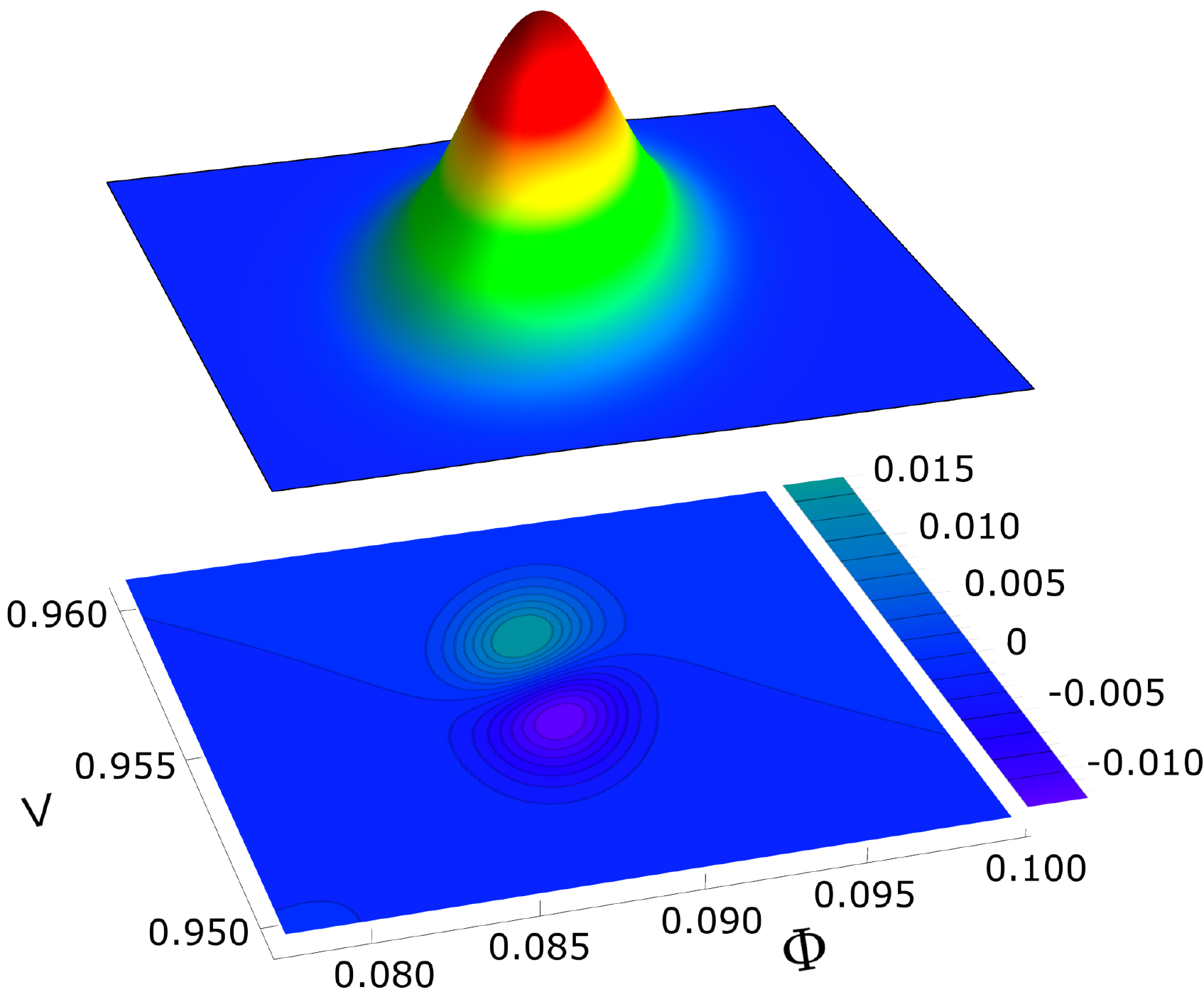}
\label{fig:fructose}

\caption{Bayesian joint probabilities for visibility ($V$) and phase ($\Phi$) (upper panel) and its difference (right-hand colored scale) with respect the one at the CRB (lower panel) for fructose (sucrose) in aqueous solution. A number $M\simeq50$K ($M\simeq75$K) of repetitions have been employed.}
\label{fig:fructose}
\end{figure*}

A more stringent test in metrology is the verification of the Cram{\'e}r-Rao bound (CRB). This sets a lower bound to the covariance matrix $\Sigma$ of the estimated parameters, whose elements are defined as the second moments of $P_B(\phi,v|\bar n)$:  
\begin{equation}
\begin{aligned}
&\Delta^2\phi = \Sigma_{\phi,\phi}=\int (\phi-\phi_B)^2 P_B(\phi,v|\bar n)d\phi\,dv,\\
&\Delta^2v = \Sigma_{v,v}=\int (v-v_B)^2 P_B(\phi,v|\bar n)d\phi\,dv,\\
&\Sigma_{\phi,v}=\Sigma_{v,\phi}=\int (\phi-\phi_B)(v-v_B) P_B(\phi,v|\bar n) d\phi\,dv.
\label{sigma}
\end{aligned}
\end{equation}
The measurement strategy is characterised by its Fisher information matrix ${\mathcal F}$, whose elements are:
\begin{equation}
\mathcal{F}_{ij}=\sum_\theta \frac{\partial_i p(\theta|\phi,v)\,\partial_j p(\theta|\phi,v)}{p(\theta|\phi,v)}
\label{eq:fisher}
\end{equation}
with $i$ and $j$ can correspond to either $\phi$ or $v$. The CRB asserts that, given a number of trials $M$, the covariance matrix is bounded as
\begin{equation}
\Sigma \geq \mathcal{F}^{-1}/M.
\label{crb}
\end{equation}
This matrix inequality holds in the asymptotic limit of a large number of trials, and sets lower bounds for the individual precisions $\Delta^2\phi$ and $\Delta^2v$, as well as on their covariance $\Sigma_{\phi,v}$. 
The conditional probabilities in~\eqref{eq:fisher}, describing our experiment, corresponds to the post-selected coincidence events of our detectors. The effect of post-selection on estimation precision has been described in detail in the literature \cite{gendra2013,combes2014}, and the consequences on our experimental results are better discussed in the Appendix. It is important to notice that, while typically post-selection has been investigated as a tool to enhance the estimation precision, in our case it is a consequence of the limitation of the experimental setup.
In Fig.~\ref{fig:calibration}  we report the measured uncertainties and covariance, in very good agreement to the expected values predicted by the CRB in \eqref{crb} with $M=\sum_\theta n_\theta$. Oscillations of the attainable precisions in \eqref{sigma} can be observed: the available information is distributed between the phase and the visibility, depending on the value of $\phi$; covariances are modulated as well, and the best estimation for either individual parameter corresponds to minimal correlation.

As an application of our protocol, we perform the estimation of the phase imparted by aqueous solutions of fructose. It is common knowledge that sugars are an interesting example of chiral molecules, able to impart a rotation to an initial linear polarization. Monitoring their optical activity via light-matter coupling can thus represent a valuable approach to infer information on their interaction with the surroundings. The most relevant environment for their application is the aqueous solution: investigation with quantum light has been undertaken in \cite{Tischlere2016} in a single-parameter approach. Fig.~\ref{fig:fructose} reports the Bayesian joint probability distribution for two different sugar aqueous solutions, namely of fructose (F) and sucrose (S) at the same nominal concentration of $c=0.3$ g/ml. The upper 3D plots show the reconstructed distributions, which give back the following average values $\phi_F=-0.145$ rad and $\phi_S=0.089$ rad, consistently with the values obtained by using classical light of a close wavelength (808 nm) in the same apparatus. The underlying contour plots show the difference between the reconstructed distribution and the expected Gaussian saturating the CRB; for both concentrations, the deviations remain of the order of 0.01. In order to assess quantitatively how close our estimation lies to the CRB, we adopt the likelihood ratio test predicting that, under the null hypothesis that $\Sigma$ saturates \eqref{crb}, the variable
\begin{equation}
l = M^2 \text{Tr}\left(F\cdot\Sigma\right)-M\left(\ln\det(\Sigma)+\ln\det(M\,F)\right)-2
\end{equation}
is distributed as $\chi^2$ variable with 3 degrees of freedom~\cite{anderson2003}. The measured values for the two concentrations are $l_{F}=2.63$, and $l_{S}=0.10$, both compatible with the critical value $7.81$ for the 95\% confidence interval. 

\section{Scaling laws for multiparameter estimation}

\begin{figure*}[t]
\includegraphics[width=0.325\textwidth]{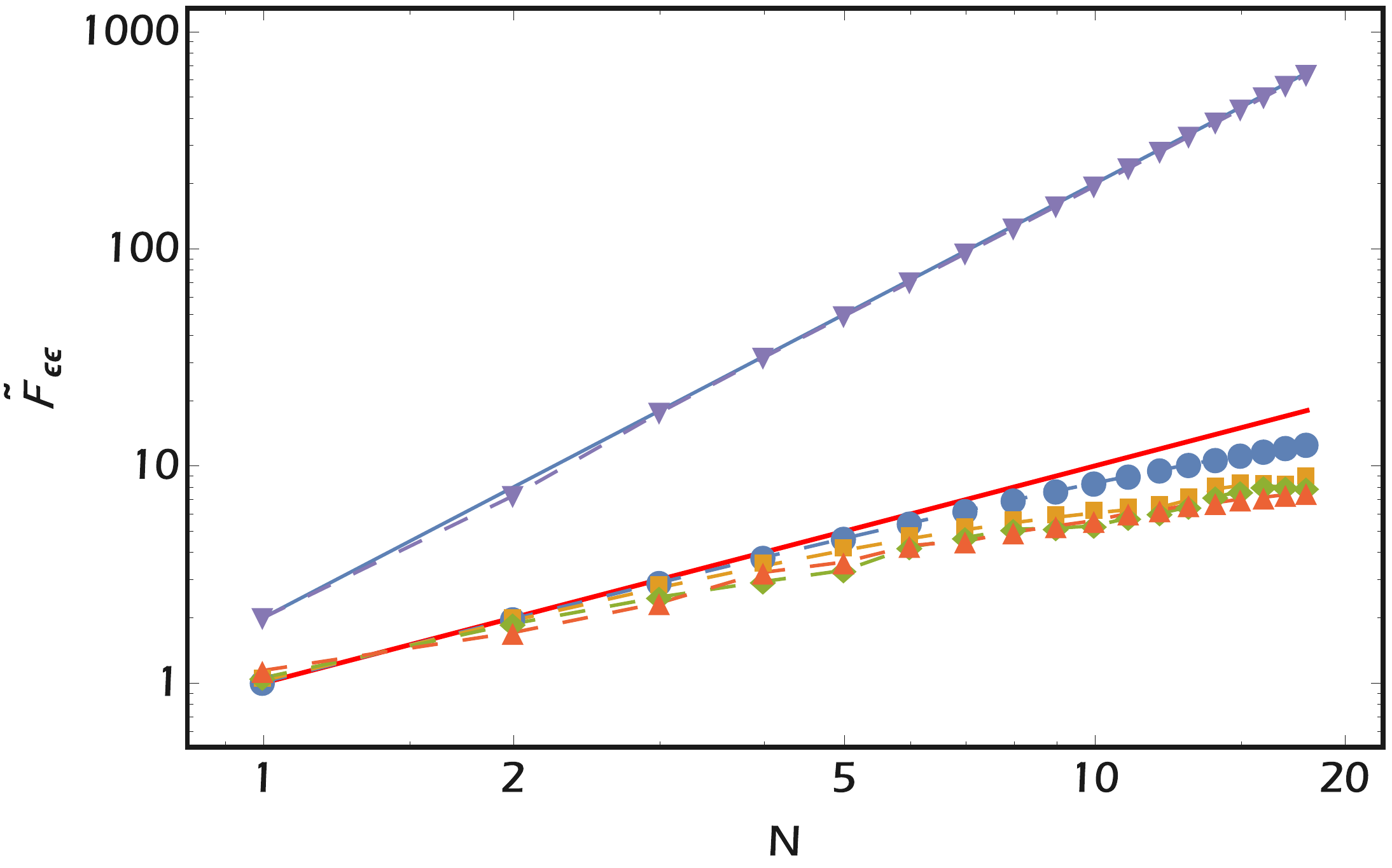}
\includegraphics[width=0.325\textwidth]{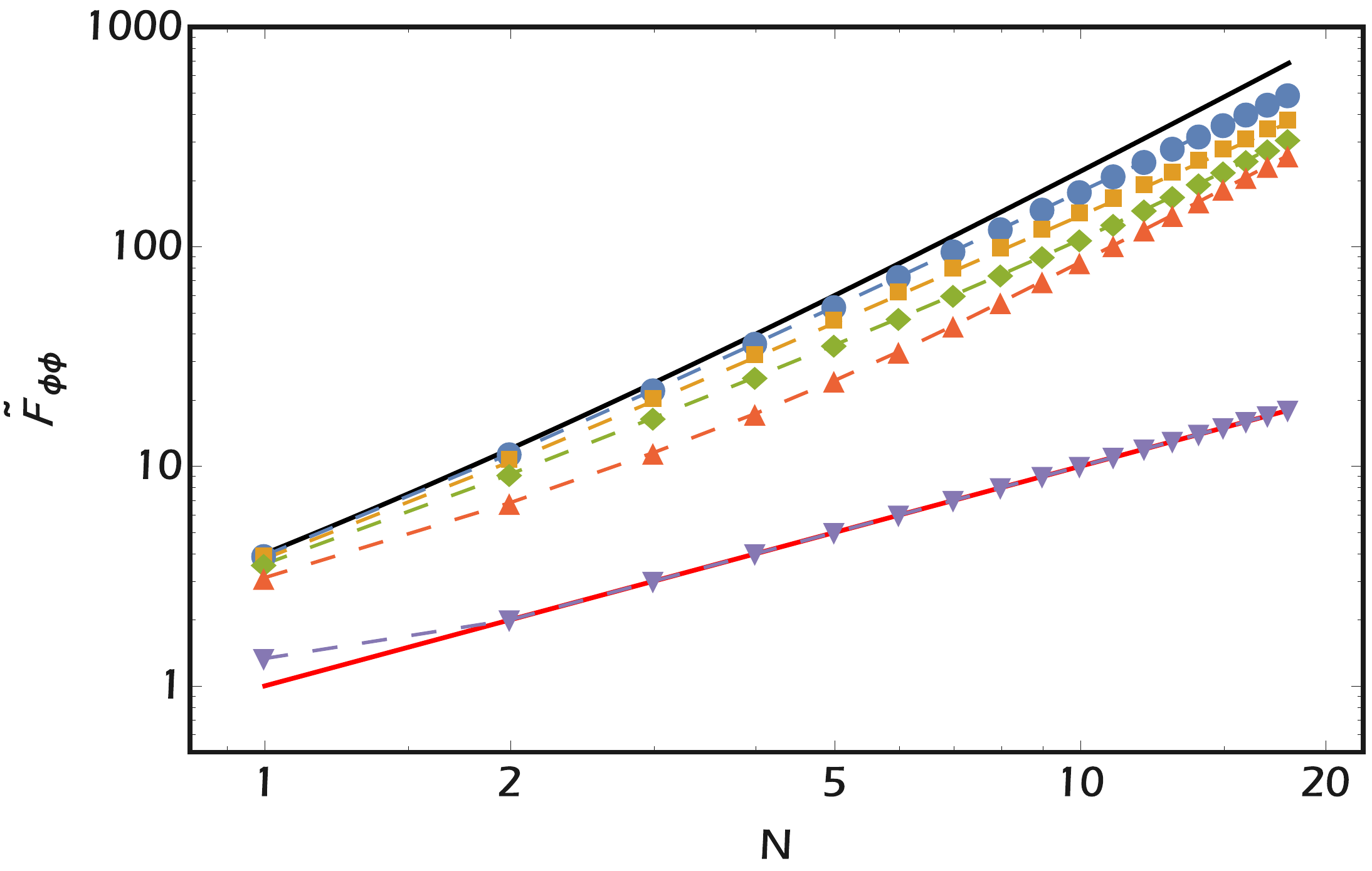}
\includegraphics[width=0.315\textwidth]{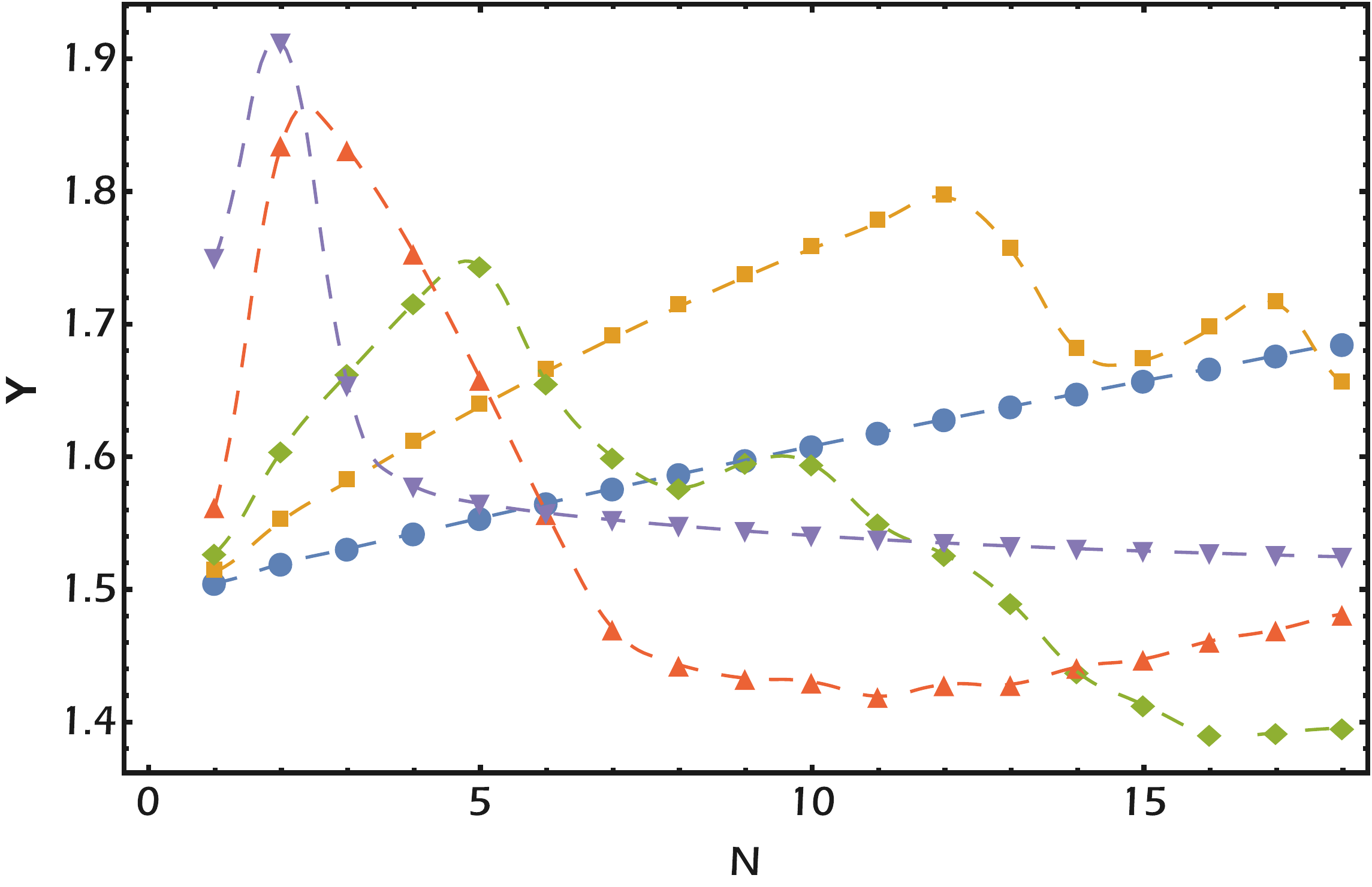}

\caption{Scaling of the Fisher information: (a) effective Fisher information for the distinguishability parameter $\epsilon$. The points correspond to numerical results, and the solid lines correspond to $2N^2$ (black) and $N$ (red). (b) effective Fisher information for the phase $\phi$. The points correspond to numerical results, and the solid lines correspond to $2N(N+1)$ (black) and $N$ (red).  (c) Trade-off in the optimality of individual estimations quantified by $\Upsilon$. In all plots: {\color{blu} $\bullet$}: $\epsilon=0.14$, {\color{ocra} $\blacksquare$}: $\epsilon=0.23$, {\color{verde} $\blacklozenge$}: $\epsilon=0.32$, {\color{rosso} $\blacktriangle$}: $\epsilon=0.50$, {\color{viola} $\blacktriangledown$}: $\epsilon=1$.}
\label{fig:scaling}
\end{figure*}

The usefulness of quantum resources is typically assessed by looking at how the precision on given parameters scale with the number of photons $N$ in the probe. for phase estimation, quantum probes can reach a scaling law o f the Fisher Information as $N^2$ while classical resources  are limited to $N$. For loss the Fisher Information grows as $N$ for both classical and quantum probes~\cite{monras2007,genoni2011}. For these purposes we generalized to states with $2N$ photons: we consider Holland-Burnett (HB) states~\cite{Holland1993} that are obtained by quantum interference of two $N$ photon states arriving on input modes with creation operators $a_{H}^\dagger$ and $b_{V}^\dagger$, which are made interfere. A phase $\phi$ is then inserted, and the detection scheme considers a second interference, followed by photon-number resolving detectors on each arms. In order to account for distinguishability, we take the standard decomposition $b_{V}^\dagger=\sqrt{1-\epsilon^2}\,a_V^\dagger+\epsilon\,q_V^\dagger$, where $a_V^\dagger$ interferes perfectly with $a_{H}^\dagger$, while $q_V^\dagger$ does not. The parameter $\epsilon$ defines the distiguishability of the two modes:  from $\epsilon=0$ for perfect indistinguishability to $\epsilon=1$ for complete distinguishability. We are interested in how the available Fisher information associated to $\phi$ and to $\epsilon$ scales with the number of photons $N$: we use as quantifiers the effective values $\tilde F_{i,i}=1/(F^{-1})_{i,i}$ for $i=\phi,\epsilon$, optimized over all possible phases.  

The results of our numerical simulations are reported in Fig.~\ref{fig:scaling}a and b. For moderate distinguishability, the effective Fisher information on $\phi$ decreases with respect to its value $2N(N+1)$ at $\epsilon=0$~\cite{Holland1993}, but retains a quicker growth than the classical scaling as $N$, obtained for $\epsilon=1$.
Nevertheless, while the effective Fisher information $\tilde{F}_{\phi\phi}$ is reduced due to the presence of correlation between the two parameters, the plot leads us to  conjecture that, as observed in \cite{Birchall2016} for the phase-estimation only problem, an asymptotic quadratic scaling is maintained also for distinguishability $0<\epsilon<1$. Regarding the distinguishability $\epsilon$ we remark a non-monotonic behavior: the information initially decreases with respect to the linear scaling, but a quadratic behaviour $2N^2$ is eventually observed in the limit $\epsilon=1$. These optimal values, however, are obtained for different phases $\phi$: in general, it is not possible to satisfy the optimality conditions for both parameters at once. In order to understand how the information is partitioned, we adopt the parameter: 
\begin{equation}
\Upsilon =\max_\gamma \left(\frac{\tilde F_{\phi,\phi}(\gamma)}{\max_\alpha \tilde F_{\phi,\phi}(\alpha)}+\frac{\tilde F_{\epsilon,\epsilon}(\gamma)}{\max_\beta \tilde F_{\epsilon,\epsilon}(\beta)}\right).
\end{equation} 

This figure of merit aims to quantify the  overall effectiveness of the measurement scheme, obtained by varying the phase $\phi$, calculating the sum of the ratios between the effective Fisher information for respectively $\phi$ and $\epsilon$, and their maximum values, reached for a particular value of $\phi$. The corresponding results are show in in Fig. \ref{fig:scaling}c: for each value of $\epsilon$ there exist a value of $N$ that achieves the best compromise in the jointly attainable precision. Our numerical results also suggest that, while for highly noisy probes (large values of $\epsilon$) the optimal value occurs for small $N$, for nearly-ideal probes (small values of $\epsilon$), optimality is reached for larger values of $N$, where the quantum enhancement in the phase estimation is more prominent.

\section{Conclusions}
Multiparameter estimation can be an effective way to tackle the problem of operating quantum sensors in the presence of noise, an unavoidable challenge in realistic conditions. We have applied such an approach to integrate phase estimation with a simultaneous characterization of the probe, by measuring phase and visibility of interference fringes at once. Depending on the value of the phase, oscillations in the achieved precision on individual parameters are observed, and correlations are introduced. The measurement scheme has been used to investigate the optical activity of fructose solutions. Numerical simulations have been undertaken to study how the precisions scale with the photon number in Holland-Burnett states.

Our results highlight the presence of trade-off conditions, as part of the information need being devoted to determine the quality of the probe at the expense of the precision on the phase. Realising the promises of quantum sensing will need to understand the price of achieving robust operation in unfavourable conditions: our study is an important step in this direction.

\section*{Acknowledgments}
We thank P. Aloe, F. Somma,  A. Sodo, F. Bruni, and M.G.A Paris for useful discussions. This work has been funded by the European Commission via the Horizon 2020 Programme (Grant Agreement No. 665148 QCUMbER). MGG acknowledges support from Marie Sk\l{}odowska-Curie Action H2020-MSCA-IF-2015 (project ConAQuMe, grant no. 701154). The Grant of Excellence Departments, MIUR (ARTICOLO 1, COMMI 314--337 LEGGE 232/2016), is gratefully acknowledged.


\bibliography{Bibliography-2}


\newpage

\begin{widetext}
\section*{Appendix}
\subsection*{Evolution of the two-photon state under phase rotation}

The aim of this section is to obtain detection the probabilities from a two-photon $N$00$N$ state with limited visibility, following a phase shift $\phi$. We start by noticing that the combination of a beam splitter (BS), a phase shift $\phi$, and a second BS can be modelled as an unbalanced BS with transmission $\cos(\phi/2)$. Since we use the polarization degree of freedom of a single spatial mode, the phase shift can be implemented as a polarization rotation by means of a half wave plate, as in the calibration phase, or of an optically active solution. The mode mixing performs the transformation:
\begin{equation}
\begin{aligned}
a_H\rightarrow \cos(\phi/2)a_H+\sin(\phi/2)a_V\\
a_V\rightarrow \cos(\phi/2)a_V-\sin(\phi/2)a_H
\label{bogo}
\end{aligned}
\end{equation}
The first step considers perfectly indistinguishable photons in the two-photon state $|\Psi_0\rangle_{in}=\hat{a}^{\dagger}_H \hat{a}^{\dagger}_V|0\rangle$, which evolve following the phase shift:
\begin{equation}
\begin{split}
|\Psi_{\phi}\rangle_{in}=\Biggr[\text{cos}\biggr(\frac{\phi}{2}\biggr) \hat{a}^{\dagger}_H +\text{sin}\biggr(\frac{\phi}{2}\biggr) \hat{a}^{\dagger}_V\Biggr]\Biggr[\text{cos}\biggr(&\frac{\phi}{2}\biggr) \hat{a}^{\dagger}_V - \text{sin}\biggr(\frac{\phi}{2}\biggr) \hat{a}^{\dagger}_H\Biggr]|0\rangle \\ 
= \Biggr[\text{cos}(\phi) \hat{a}^{\dagger}_H \hat{a}^{\dagger}_V - \text{sin}(\phi) \frac{(\hat{a}^{\dagger^2}_H - \hat{a}^{\dagger^2}_V)}{2}\Biggr]|0\rangle.
\end{split}
\label{instate}
\end{equation}
A half wave plate (HWP) is inserted, and set at an angle $\theta$ with respect to the horizontal; its effect on the two-photon state delivers the expression
\begin{equation}
|\Psi_{\phi;\theta}\rangle_{in}=\text{cos}(\phi) \Biggr[-\text{cos}(4\theta)\hat{a}^{\dagger}_H \hat{a}^{\dagger}_V + \frac{1}{2}\text{sin}(4\theta)(\hat{a}^{\dagger^2}_H - \hat{a}^{\dagger^2}_V)\Biggr]|0\rangle - \text{sin}(\phi)\Biggr[\text{sin}(4\theta)\hat{a}^{\dagger}_H \hat{a}^{\dagger}_V + \frac{1}{2}\text{cos}(4\theta)(\hat{a}^{\dagger^2}_H - \hat{a}^{\dagger^2}_V)\Biggr]|0\rangle,
\end{equation}
from which the following probabilities can be obtained:
\begin{equation}
\begin{split}
p^{in}_1(\theta|\phi)&=\frac{1}{2}(1+\text{cos}(8\theta - 2\phi))\\
p^{in}_2(\theta|\phi)&=\frac{1}{2} \text{sin}^2(4\theta - \phi).
\end{split}
\end{equation}

Using the same approach, it is possible to calculate the evolution of quantum state $|\Psi_0\rangle_{dis}=\hat{a}^{\dagger}_H \hat{b}^{\dagger}_V|0\rangle$ for two distinguishable photons: here we also need considering two extra modes $a_V$ and $b_H$, initially in the vacuum mode, to define transformations similar to the ones in \eqref{bogo}. These then give an expression for the state as:
\begin{equation}
\begin{split}
|\Psi_{\phi}\rangle_{dis}=\Biggr[\text{cos}\biggr(\frac{\phi}{2}\biggr) \hat{a}^{\dagger}_H + \text{sin}\biggr(\frac{\phi}{2}\biggr) \hat{a}^{\dagger}_V\Biggr] \Biggr[\text{cos}\biggr(\frac{\phi}{2}\biggr) \hat{b}^{\dagger}_V - \text{sin}\biggr(\frac{\phi}{2}\biggr) \hat{b}^{\dagger}_H\Biggr]|0\rangle\\ 
= \Biggr[\text{cos}\biggr(\frac{\phi}{2}\biggr)^2 \hat{a}^{\dagger}_H \hat{b}^{\dagger}_V -\text{sin}\biggr(\frac{\phi}{2}\biggr)^2 \hat{a}^{\dagger}_V \hat{b}^{\dagger}_H - \text{sin}(\phi) \frac{(\hat{a}^{\dagger}_H \hat{b}^{\dagger}_H - \hat{a}^{\dagger}_V \hat{b}^{\dagger}_V)}{2}\Biggr]|0\rangle.
\end{split}
\label{eqout}
\end{equation}
and then, including the HWP, the probabilities:
\begin{equation}
\begin{split}
p^{dis}_1(\theta|\phi)&=\frac{1}{4}(3+\text{cos}(8\theta - 2\phi))\\
p^{dis}_2(\theta|\phi)&=\frac{1}{4} \text{sin}^2(4\theta - \phi).
\end{split}
\end{equation} 
These have been obtained considering detectors unable to distinguish between the modes $a$ and $b$. 

In the general case, the initial mode $b_V$ will possess a component $a_V$ indistinguishable from $a_H$ in all other degrees of freedom, and a distinguishable component $q_V$: $b_V^\dagger=\sqrt{1-\epsilon^2}a_V^\dagger+\epsilon q_V^\dagger$. As before, we need to introduce extra vacuum modes. The final probabilities will be given by the weighted sums:
\begin{equation}
\begin{split}
p_1(\theta|\phi,\epsilon)&=\frac{1-\epsilon^2}{2}(1+\text{cos}(8\theta - 2\phi))+\frac{\epsilon^2}{4}(3+\text{cos}(8\theta - 2\phi))\\
p_2(\theta|\phi,\epsilon)&=\frac{1-\epsilon^2}{2} \text{sin}^2(4\theta - \phi)+\frac{\epsilon^2}{4} \text{sin}^2(4\theta - \phi).
\end{split}
\end{equation}
These expressions can be cast in more compact form by introducing the visibility $v$ of the predicted fringes as $v=(2-\epsilon^2)/(2+\epsilon^2)$: 
\begin{equation}
\begin{split}
p_1(\theta|\phi,v)&=\frac{1}{1+v}\left(1+v\cos(8\theta - 2\phi)\right)\\
p_2(\theta|\phi,v)&=\frac{v}{1+v} \sin^2(4\theta - \phi).
\label{POVM}
\end{split}
\end{equation}

\subsection*{Optimal measurement for single parameter estimation approach}
We now investigate the performance of phase estimation using a single-parameter approach that relies on a pre-calibration of the visibility $v_0$. The post-selected probabilities are given by
\begin{equation}
p(\theta|\phi)=\frac{1}{4}(1+v_0\cos(8\theta - 2\phi)),
\end{equation}
where the error on $v_0$ is considered negligible. The ultimate bound on the phase precision is than given by $1/M\mathcal{F}_{\phi\phi}$: we compare the uncertainties on the phase to this limit, considering three instances for the visibility, namely the maximum, the mean and the minimum values obtained in the multiparameter analysis. By inspecting Fig.~\ref{1par} it appears evident how the extreme values  $v_{min}$ or $v_{max}$ lead to an estimation that significantly departs from its CRB, even showing a violation thereof. The use of the mean visibility mitigates such a discrepancy, although it manifests differences with respect to the CRB three times larger than those achieved with the multi-parameter approach.

\begin{figure}[ht!]
\begin{center}
\includegraphics[width=9 cm]{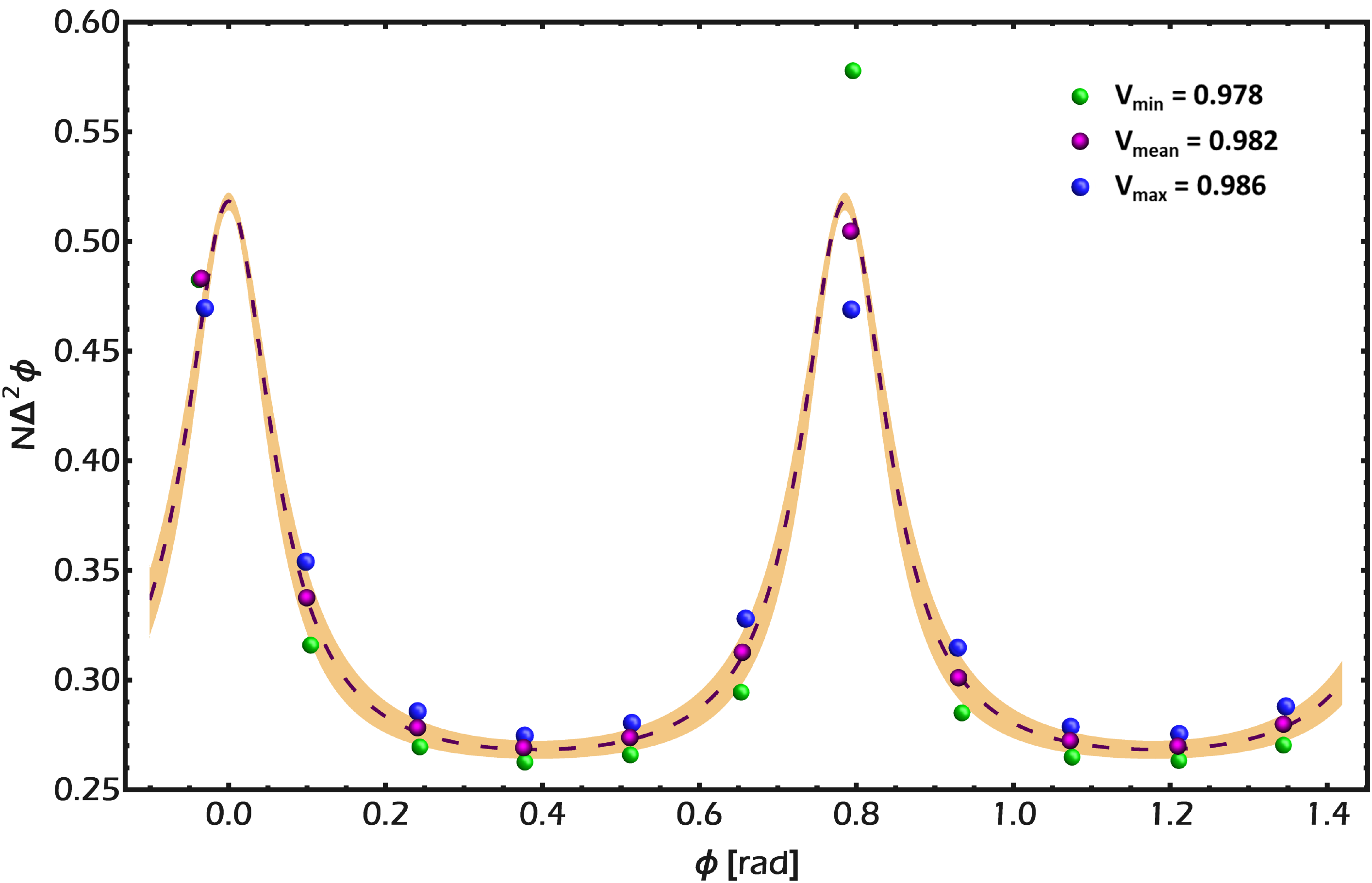}
\end{center}
\caption{The estimated phase variance for three values of visibility, $v_{\text{min}}=0.978$ (green dots), $v_{\text{mean}}=0.982$ (purple dots), $v_{\text{max}}=0.986$ (blue dots), compared with the expected CRB for pre-calibrated visibilities ranging from $v_0=v_{\text{min}}$ to $v_0=v_{\text{max}}$ (shaded area): $v_0=v_{\text{mean}}$ is highlighted (purple dashed line).}
\label{1par}
\end{figure}

\subsection*{Effects of the post-selection on the Fisher information}
The set \eqref{POVM} describe the probabilities of a three-outcome POVM for any given $\theta$. We can generalise this to a set of $m \times 3$ outcomes by choosing $m$ different values of $\theta$ with different probabilities $p(\theta)$; the relevant case for our experiment has $m=4$ with $\theta=\{0,\;\pi/16,\;\pi/8,\; 3\pi/16\}$, and a flat distribution for the settings $p(\theta)=1/4$. The Fisher information matrix associated to this strategy is found as:
\begin{equation}
F_{ij}=\frac{1}{4}\sum_{\theta} \left( \frac{\partial_i p_1(\theta|\phi,v) \partial_j p_1(\theta|\phi,v)} {p_1(\theta|\phi,v)}+2 \frac{\partial_i p_2(\theta|\phi,v) \partial_j p_2(\theta|\phi,v) } {p_2(\theta|\phi,v)} \right),
\end{equation} 
for $i=\phi,v$ and likewise for $j$. We remark that with photon number resolving detectors, a POVM with $m=2$ settings would be able to provide the same Fisher information.

In the actual experiment, we have no access to the full set of the outcomes, since we did not use photon number resolving detectors. We then use the post-selected probabilities 
\begin{equation}
\begin{split}
p(\theta|\phi,v)&=\frac{1}{4}\left(1+v\cos(8\theta - 2\phi)\right)
\label{PS}
\end{split}
\end{equation}
associated to coincidence counts only, and obtained by normalising each of the four $p_1(\theta|\phi,v)$ to their sum. As described in the main text, in the post-selected picture, the setting $\theta$ plays the role of the measurement outcome: given the coincidence event \eqref{PS} quantifies the probability that this has occurred by setting the HWP angle at the value $\theta$. The post-selected Fisher information is then given by:
\begin{equation}
F^{PS}_{ij}=\frac{1}{4}\sum_{\theta} \frac{\partial_i p(\theta|\phi,v) \partial_j p(\theta|\phi,v)} {p(\theta|\phi,v;\theta)},
\end{equation} 
which is the one used in the main text. We notice that adopting this strategy results in a loss of the useful resources by a factor 2. 

A comparison of the two Fisher matrices is carried out in Fig.~\ref{comp}: post-selection affects phase estimation by reducing the available information $F_{\phi,\phi}$, while the visibility estimation appears improved by a higher value of $F_{v,v}$. We notice that this increase, however, does not compensate the loss of resources: if one takes into account the probability of the favourable events that are post-selected, the weighted post-selected Fisher Information is in general always lower than the Fisher information of the complete POVM \cite{combes2014}. Furthermore, the correlation properties are also made tighter: this is verified by introducing a normalised value for the off-diagonal terms as $\xi_{\phi,v}=F_{\phi,v}/(F_{\phi,\phi} F_{v,v})^{1/2}$ showing more pronounced oscillations. Notice that no spurious correlations are introduced when these are absent in the original POVM.  

\begin{figure*}[h!]
{\includegraphics[width=0.33\textwidth]{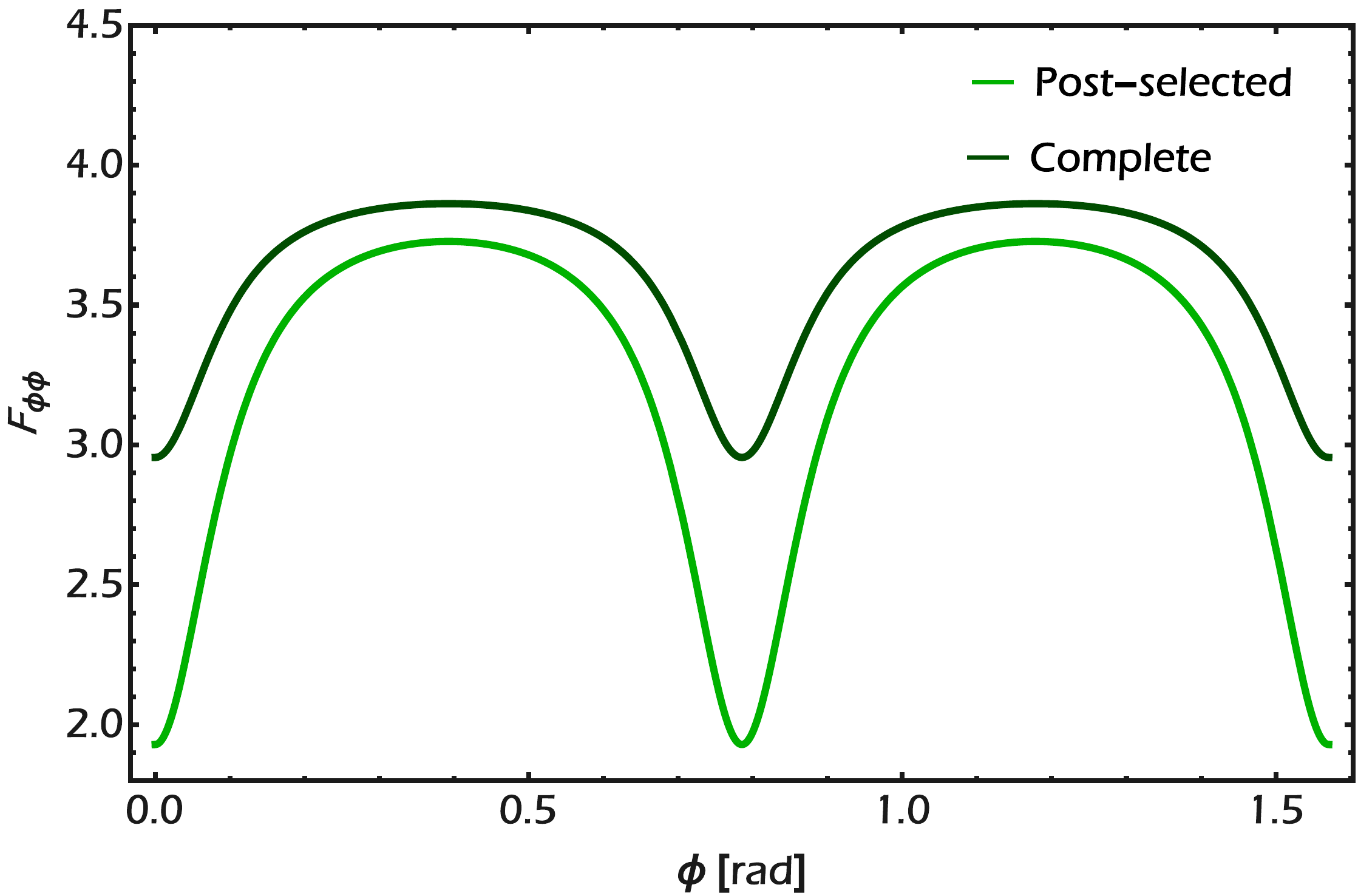}}
{\includegraphics[width=0.33\textwidth]{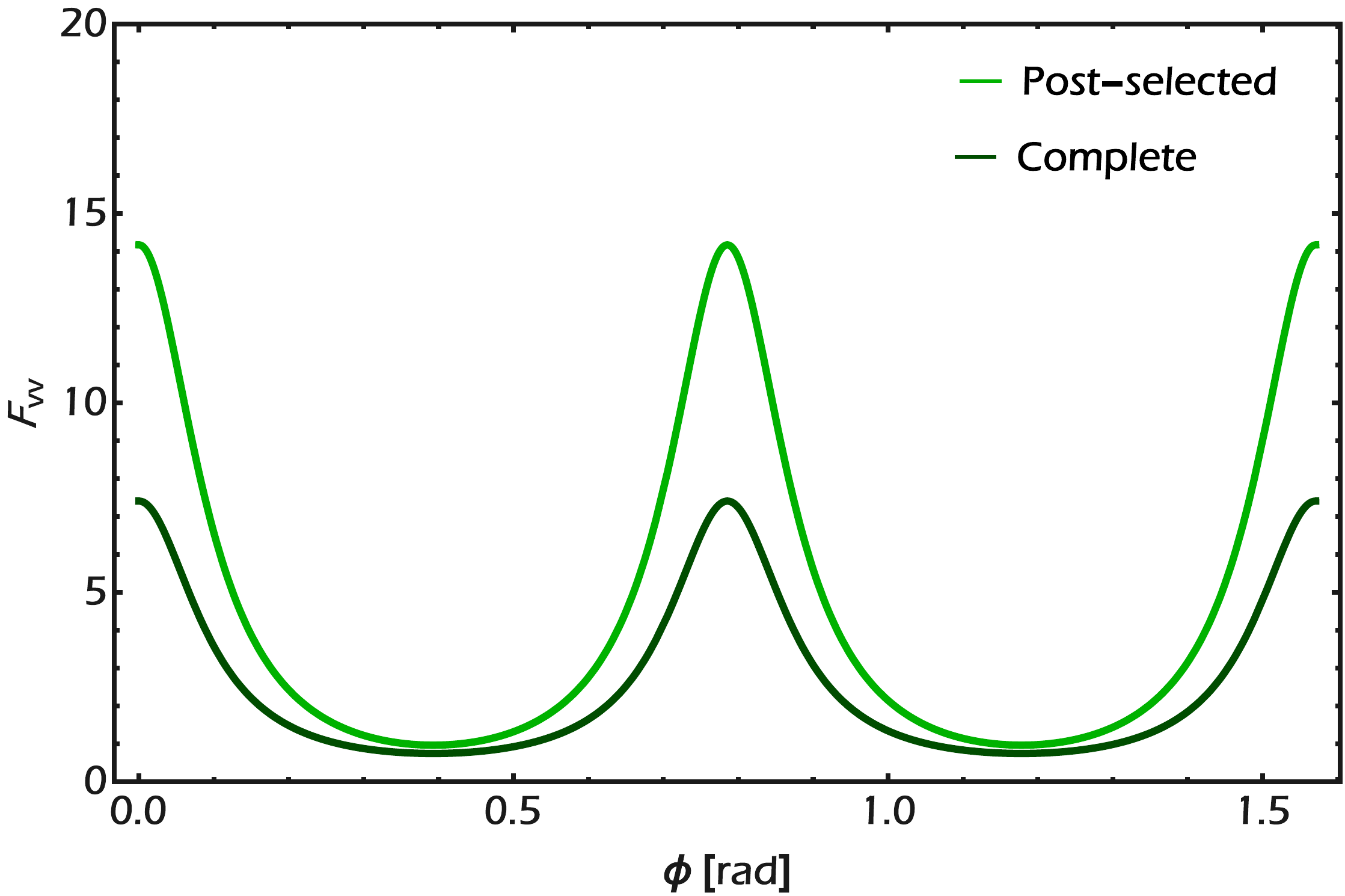}}
{\includegraphics[width=0.33\textwidth]{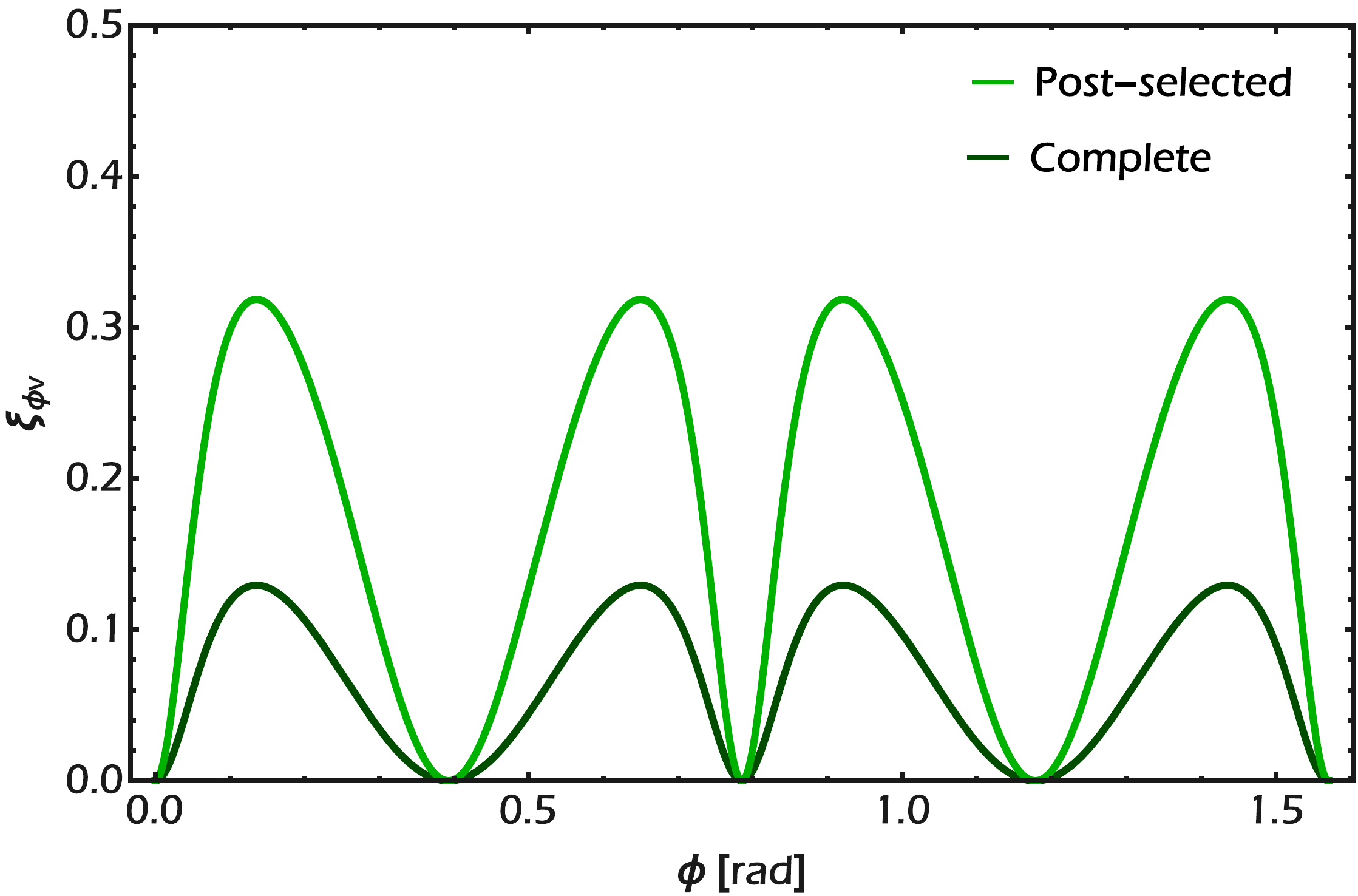}}
\caption{The post selected (bright green line) and complete (dark green line) Fisher information matrix elements, $F_{\phi,\phi}$ (a), $F_{v,v}$ (b) and $\xi_{\phi,v}$ (c).}
\label{comp}
\end{figure*} 

\section{Generalization to 2$N$-photon Holland Burnett states}
We now generalize our approach to arbitrary Holland-Burnett states, obtained by the interference of two $N$-photon Fock states. The initial state is 
\begin{equation}
|\Psi_0\rangle=\frac{1}{N!}(a_H^\dagger)^N (b_V^\dagger)^N \vert 0\rangle = \frac{1}{N!}(a_H^\dagger)^N (q_H^\dagger)^0 \sum_{k=0}^{N} \binom{N}{k} (1-\epsilon^2)^{k/2}\epsilon^{(N-k)/2} (a_V^\dagger)^k (a_V^\dagger)^{N-k} \vert 0\rangle. 
\end{equation}
Here we have operated the same modal decomposition as before to introduce distinguishability, and we have made the presence of the extra modes explicit. If the transformations \eqref{bogo} are imposed on the pairs of modes $a_H$ and $a_V$, and $q_H$ and $q_V$, the evolved state $|\Psi_\phi\rangle$ can be calculated. In the general case, an additional controlled phase $\theta$ can be introduced, which corresponds to different measurement settings.

The POVM we consider counts the total photon number on modes $a_H$ and $q_H$, without resolving the individual populations; due to photon number correlations, adding a second counter on the modes 
$a_V$ and $q_V$ would provide no extra information. The operator associated to the outcome $x$ is then written in the Fock basis as:
\begin{equation}
\Pi_{x} =\sum_{s=0}^x \Pi^{(s)}_{x}=\sum_{s=0}^x \vert{s}\rangle\langle s\vert_{aH}  \otimes I_{aV}  \otimes \vert{x-s}\rangle\langle x- s\vert_{qH} \otimes I_{qV}. 
\end{equation}
Each detection probability is then found as $p(x,\theta|\phi,\epsilon) =\langle \Psi_{\phi+\theta}\vert \Pi_x \vert\Psi_{\phi+\theta}\rangle = \sum_{s=0}^x p^{(s)}(x,\theta|\phi,\epsilon)$ for all possible outcomes $x=0,...,2N$. In our calculations we considered two possible settings $\theta=0$ and $\theta=\pi/2$, alternated with equal probability.  

\end{widetext}

\end{document}